**IAC-25-A4.2.4 (99704)**

# SETI Post-Detection Protocols: Progress Towards a New Version

**Michael A. Garrett*[a], Kathryn Denning[b], Leslie I. Tennen[c], Carol Oliver[d]**

[a] University of Manchester, Jodrell Bank Centre for Astrophysics, UK, and University of Leiden, Leiden Observatory, NL. michael.garrett@manchester.ac.uk
[b] Dept of Anthropology, York University, Toronto, Canada. kdenning@yorku.ca
[c] Law Offices of Sterns and Tennen, Phoenix, Arizona, USA. ltennen@astrolaw.com
[d] Australian Centre for Astrobiology, School of Biological, Earth, and Environmental Sciences, University of New South Wales, Australia. carol.oliver@unsw.edu.au

* Corresponding Author

**Abstract**

The International Academy of Astronautics (IAA) SETI Committee has long provided guiding principles for responding to a potential detection of a SETI signal. The foundational *Declaration of Principles Concerning Activities Following the Detection of Extraterrestrial Intelligence*, first formulated in 1989, has been widely recognised by the international scientific community. A supplemental set of draft protocols addressing the possibility of a reply to an extraterrestrial signal was prepared in 1995 by the IAA SETI Permanent Committee, with both documents presented in a position paper to the UN Committee on the Peaceful Uses of Outer Space in 2000.

In keeping with the evolving landscape of SETI research, the IAA *Declaration of Principles* was streamlined and updated in 2010. Recognising the need for continued adaptation, the IAA SETI Committee established a Task Group in 2022 to re-examine the protocols in light of recent advances in search methodologies, the expansion of international participation in SETI, and the increasing complexity of the global information environment. The Group recognises the living document nature of the protocols, which will require ongoing refinement to remain relevant and effective in a rapidly changing world.

A preliminary report was presented at the 2023 International Astronautical Congress (IAC) in Baku, outlining proposed revisions. A draft revised *Declaration of Principles* was presented at the IAC 2024 in Milan, and initial feedback was received from the community, particularly members of the IAA SETI Committee. Since then, we have continued to seek broader community input in a structured process, refining the proposed updates based on further discussions and consultations.

At the IAC 2025 in Sydney, we intend to present the outcomes of this process, including a new draft of the *Revised Declaration of Principles*, which will also be submitted to the IAA SETI Committee at its annual meeting during the Congress.

## 1. Introduction

The International Academy of Astronautics (IAA) SETI Committee has played a central role in developing guiding principles for the scientific community's conduct in the event of the possible detection of extraterrestrial intelligence (ETI) [1, 2]. The original *Declaration of Principles Concerning Activities Following the Detection of Extraterrestrial Intelligence*, drafted in the 1980s and adopted in 1989, established a voluntary framework for best practices in verification, information sharing, and international consultation [3]. It was supplemented in 1995 by the *Draft Declaration of Principles Concerning the Sending of Communications to Extraterrestrial Intelligence* addressing the possibility of responding to an extraterrestrial signal [4,5,6]. Both documents were presented to the UN Committee on the Peaceful Uses of Outer Space (UNCOPUOS) in 2000 [7,8], receiving broad recognition from the international scientific community.

A streamlined update, *The Declaration of Principles Concerning the Conduct of the Search for Extraterrestrial Intelligence*, was adopted by the IAA SETI Committee in 2010 [9]. The authors of the 2010 document retained the core aims of the 1989 document and primarily framed it around radio SETI. It pre-dated the current, fast-moving communications landscape shaped by the internet, social media and Artificial Intelligence (AI). Since then, significant developments in SETI science, including the diversification of technosignature search methodologies and the growth of a global, inter-disciplinary research community, have underscored the need for further revision.

Following earlier work in 2018 and 2019, in 2022 the IAA SETI Committee established a Post-Detection Protocol Task Group to prepare an updated version of the 2010 *Declaration of Principles Concerning the Conduct of the Search for Extraterrestrial Intelligence*, with the aim of preserving the enduring values of earlier documents while making them relevant to the realities of twenty-first century science, media, and public engagement. This process was first reported at the 74th International Astronautical Congress (IAC) in Baku in 2023 [10] and a *Draft Revised Declaration* was presented at the 75th IAC in Milan in 2024 [11]. The Milan draft incorporated community consultation and addressed issues such as broader definitions of technosignatures, the role of social media, and the re-establishment of a dedicated Post-Detection Subcommittee.

Since Milan, the Task Group has sought further structured feedback from both within the IAA SETI Committee and the wider SETI community, to present this broadly supported revision to the Committee at the 76th IAC in Sydney in 2025.

**2. Context**

The revision process has unfolded against a backdrop of growing international interest in post-detection issues. Developments include:

- A significant interest within the NASA Astrobiology Program concerning standards of evidence for biosignatures and potential discoveries of life beyond Earth, and the communication of such science results to public audiences. For example, NASA Astrobiology's international workshop *Communicating Discoveries in the Search for Life in the Universe* (CDSLU) in spring 2024 brought scientists and communication researchers and practitioners together to consider challenges in public communication and ways to meet them [12]. The NASA Decadal Astrobiology Research and Exploration Strategy [13] process underway in 2024-2026 has also explicitly called for research community input on the topic of preparing for discovery, and astrobiology in society. These conversations continue to involve technosignatures researchers and IAA SETI Committee members.
- The SETI Post-Detection Hub at the University of St Andrews, founded in late 2022, which brings international, interdisciplinary researchers together to explore post-detection questions including those of societal impact and governance.
- The 2024 launch of the International Institute of Space Law's Working Group on SETI and Law, which addresses the legal and regulatory dimensions of possible post-detection scenarios.

The IAA *Declaration of Principles Concerning Activities Following the Detection of Extraterrestrial Intelligence* of 1989 was an important early precedent in defining scientists' responsibilities and best practices for communicating about discoveries of life beyond Earth with a global audience. Astrobiology science is now wrestling with similar complex issues for biosignatures in NASA-led efforts, and both technosignatures and biosignatures researchers must contend with the increasingly complex social media landscape and the realities of new space actors. The IAA SETI Committee's decision to further revise the 2010 *Declaration of Principles* has been undertaken bearing in mind the need for best practices in SETI to be shaped, at least in part, by those directly engaged in technosignatures science, while also drawing on the valuable expertise of related disciplines — including astrobiology, anthropology, law, the social sciences, communication, and policy. This approach reflects the SETI community's commitment to ensuring that the updated *Declaration of Principles* is founded on scientific rigour, while remaining integrated within the broader framework of post-detection planning.

**3. The Process**

At the IAC 2024 SETI & Society Session (A4.2) in Milan, we presented an update on our efforts to develop a revised version of the 2010 *Declaration of Principles Concerning the Conduct of the Search for Extraterrestrial Intelligence* [14]. The 2024 *Draft Revised Declaration* was formally introduced to the IAA SETI Committee (SC) during its annual meeting on 16 October 2024. Initial feedback from the Committee and others was broadly positive.

In spring 2025, we began the next phase of consultations. To facilitate structured feedback from the Committee, the IAA SC Post-Detection Protocol Task Group created a Google Forms document, enabling members to provide targeted comments on specific Principles within the new Draft. The form allowed the respondent to rate their satisfaction with each Principle on a scale from 1 to 10, and to add written comments. In some instances, such as the definition of technosignatures, we explicitly requested suggestions for improved wording. The use of Google Forms was also intended, in part, to ensure the anonymity of respondents, although members also had the option of submitting comments directly via email with their names included. The Task Group created a separate demographic form to encourage respondents to help us better understand the reach and inclusivity of the feedback process while maintaining their anonymity.

At the end of April 2025, we sent a covering email to members including the October 2024 *Draft Revised Declaration*, links to the Google Forms feedback forms, and additional background material from the IAA SETI Committee. These resources included the 2010 *Declaration of Principles*, a document outlining the differences between the October 2024 and 2010 versions, and our earlier IAC papers describing the revision process. Members were given three weeks to respond, with a deadline of 19 May 2025.

Approximately 15% of the membership provided feedback. The Task Group organised and summarised the feedback by Principle, with the assistance of ChatGPT, enabling more effective identification of where revisions were needed. We then produced a new Draft Revised Declaration, taking the feedback into account. The IAA SC Executive and the Protocols Task Group agreed on the new draft version on 1 August 2025.

Compared to the 2010 document, the August 2025 *Draft Revised Declaration of Principles* incorporated several substantive changes:

- **Endorsement structure** – A key issue raised was whether organisations should formally endorse the *Declaration* as signatories. The Task Group determined that the protocols should be regarded as the collective responsibility of the international SETI community, rather than documents to be signed by institutions, organisations, or individuals. The intention is for the document to represent best practice endorsed collectively by the IAA SETI Committee membership and the broader SETI community.
- **Broadened scope** – expansion from a primary focus on radio signals to encompass technosignatures more generally.
- **Expanded Preamble** – addition of new considerations, including the development of supplementary Best Practices and Codes of Conduct, and clarification of the respective responsibilities of individuals and organisations in public communication.
- **Post-detection research and ethics** – enhanced provisions in Principle #7 relating to post-detection research, and new language in Principle #9 addressing ethical and legal considerations.
- **Revised response clause** – an updated Principle #8 on responses to confirmed detections. The October 2024 draft had used stronger wording than the 2010 version; the 1 Aug 2025 draft presented alternative less stringent wording for discussion. The Task Group anticipates further debate as details of the changes remain under discussion.

The Task Group then circulated the August 2025 *Draft Revised Declaration of Principles* to the IAA SETI Committee **plus** the wider SETI research community for anonymous feedback, using a slightly modified version of the previous Google Forms survey asking for evaluation and commentary on the Draft's content and wording, and a separate demographics form. The community was given three weeks to respond, with a deadline of 26 August 2025. The invitation for comment was sent directly to:

- The IAA SETI Committee members email list
- The IAA SETI Community email list, which reaches a substantial proportion of researchers active in this field and includes ~215 people who are not Committee members
- The Order of the Octopus, a network of early career researchers in SETI with biannual in-person meetings and an online communication forum including ~120 people (https://theorderoftheoctopus.org/ )
- The SETI Post-Detection Hub at the University of St. Andrews, mentioned above, (https://seti.wp.st-andrews.ac.uk/), which in August 2025 included ~50 people
- Approximately 25 experts in space law including members of the International Institute of Space Law SETI & Law Working Group

Some of these lists overlap: for example, some Hub members are also IAA SETI Committee members, and some OOTO members are also on the IAA SETI Community list, etc. Nonetheless, in total, the invitation was sent to over 350 individuals, thereby substantially expanding the opportunity for comment beyond the IAA SETI Committee.

The feedback received from the community was, overall, strongly positive. The Task Group received 30 responses to the main form concerning the content and wording of the Aug 2025 *Draft Revised Declaration*. Figure 1 (below) presents how satisfied respondents to the feedback form were with respect to the August 2025 draft – over 70% of respondents rated their satisfaction with the document at 8 or higher on a 10-point scale.

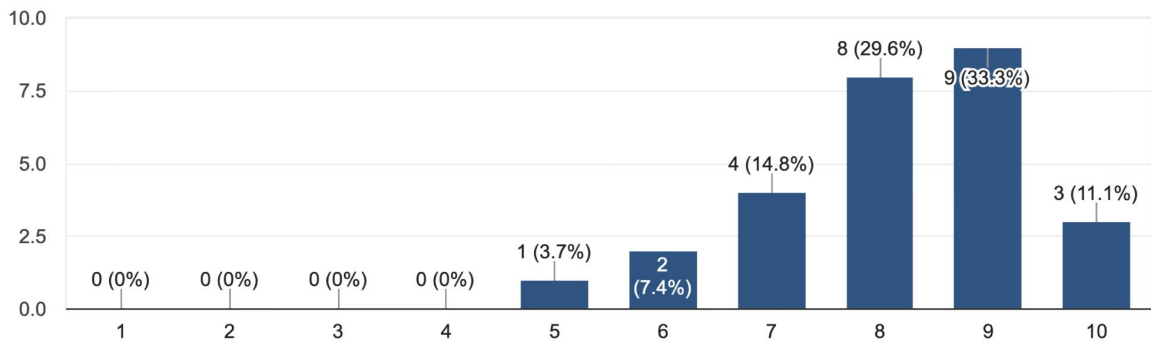

**Figure 1:** Respondents' rating of their general satisfaction with the August 2025 *Draft Revised Declaration of Principles*

The Task Group received 24 responses to the associated optional demographics form, i.e. most respondents also voluntarily provided additional information about their age range, gender, professional expertise, and geographical region. This allowed us to see that overall, we received feedback from individuals in a balanced distribution of age categories and career stages, gender, from a range of expertise ( astronomy, physics, software, engineering, data science, natural science, social science, law/ policy, humanities and creative arts, and communications), and from multiple continents.

The detailed feedback concerning the content and wording of the *Declaration* was incorporated into the further revised version 25 September 2025 for presentation to the IAA SETI Committee membership in the Committee meeting at the IAC 2025 on 1 October. This may be found in Appendix A: Draft Revised *Declaration of Principles Concerning the Conduct of the Search for Extraterrestrial Intelligence (SETI) – 2026 Update.*

Compared with the August 2025 draft, the September 2025 version incorporates several changes. Principles 4 and 5 have been consolidated. Other changes include: a clearer articulation of procedures for handling candidate evidence; greater emphasis on transparency in communication with both the scientific community and the public; and strengthened provisions on long-term data archiving and accessibility. Additional language has been inserted to highlight the importance of protecting the safety of researchers, and to expressly mention risk communication as well as science communication. Finally, communications with ETI are addressed: SETI practitioners should work with the UN and other appropriate international bodies to consider any reply to a confirmed detection. Pending the outcome of such consultations, no reply should be sent.

The input from the wider community prompted useful reflection on the scope and remit of the Committee. While some suggestions were deemed to lie beyond the current mandate of the IAA SETI Committee, the revised draft nonetheless acknowledges the importance of future work in a number of different areas.

### 4. Next steps

The next steps are to present the 25 September 2025 draft *Declaration of Principles Concerning the Conduct of the Search for Extraterrestrial Intelligence (SETI) – 2026 Update* to the IAA SC membership at their annual meeting during IAC 2025. This will represent another opportunity for members to provide comments and propose changes to the document. Hopefully, after extensive consultation, any proposed changes will now be minor. If the draft receives a favourable reception in Sydney, the final version (or possibly a subsequent minor revision) will be circulated to the IAA SETI Committee by the Chair, who will request a formal vote from the membership on its ratification and adoption. If all goes according to schedule, voting will occur before the end of 2025. The vote will use a secure, anonymous online voting system to ensure both confidentiality and integrity of the outcome. A simple majority vote will adopt the new *Declaration of Principles Concerning the Conduct of the Search for Extraterrestrial Intelligence (SETI) – 2026 Update.*

The final step in this process will be to submit the approved version of the revised *Declaration of Principles* to the IAA Board of Trustees for their endorsement. Other bodies within the IAA, such as the Space Physical Sciences Commission, will also be informed of progress. Additionally, other relevant organizations will be notified. The SETI Committee Executive anticipates that the document can be submitted to the IAA Board of Trustees for endorsement in the first half of 2026, paving the way for the formal presentation of the newly ratified protocols at IAC 2026 in Antalya, Türkiye. Given the significant public and media interest in the Protocols, this venue will provide an excellent opportunity to present the work to a broad-based international audience. The final adoption of the new document will represent a significant milestone for the global SETI community in establishing updated, consensus-driven best-practice principles for the conduct of the Search.

### 5. Conclusions

The revision of the IAA SETI Post-Detection Protocols is a multi-year process, informed by historical precedent, evolving scientific practice, and extensive community consultation. The current draft reflects a broadened scope to encompass technosignatures more generally, the inclusion of new provisions on post-detection research and ethics, and an expanded Preamble setting the stage for supplementary Best Practices and Codes of Conduct. While some elements, notably the language of Principle #8 concerning responses to confirmed detections, remain under discussion, the draft embodies a clear consensus on the need for principles that are both scientifically rigorous and adaptable to the realities of the contemporary research and communications environment. The latest draft Declaration of Principle is grounded in technical expertise while also being responsive to legal, social, and policy considerations. The active participation of the wider SETI community and allied organisations has been critical in shaping a document that aspires to serve as a durable yet flexible foundation for best practice in the decades ahead.

The conclusion of the current process is now in sight, although there is still much future work to be done on Best Practices, Codes of Conduct, and procedures for international consultations regarding reply transmissions. Once the revised *Declaration* has been adopted by the IAA SETI Committee and endorsed by the IAA Board of Trustees, it will provide an updated framework for the responsible conduct of technosignature research. Its implementation, together with the development of supplementary guidelines, and the reconstitution of a dedicated Post-Detection Subcommittee, will strengthen the community's ability to respond in a coordinated, transparent, and scientifically rigorous manner should credible evidence of extraterrestrial intelligence be discovered.

*Appendix A*

**DRAFT**

**DRAFT REVISED Declaration of Principles Concerning the Conduct of the Search for Extraterrestrial Intelligence (SETI) – 2026 Update**
**(version: 25 September 2025)**

**Preamble**

The International Academy of Astronautics SETI Committee has adopted these Principles to guide individuals, institutions, organisations, and other entities participating in the scientific Search for Extraterrestrial Intelligence (SETI), that is, the search based on astronomy and related disciplines for 'technosignatures'[1] or evidence of past or present intelligent life and technology beyond Earth.

The purpose of this Declaration is to affirm our commitment to conduct the Search for Extraterrestrial Intelligence in a scientifically and academically rigorous manner; to establish best practices, principles and guidelines for scientists to confirm putative evidence of intelligent extraterrestrial beings; to provide guidance to the scientific community for the announcement of a confirmed SETI detection which balances the community imperative of providing timely and accurate information to a wide-ranging audience, with appropriate consideration for the safety and exposure of individual scientists involved; and to proactively inform the global public of these procedures and guidelines.

The commitments in this Declaration are made with the recognition that the scientific and communications landscapes are ever-changing, and this Declaration will be supplemented with Best Practices Guidelines, including guidance concerning safety for researchers, and a Code of Conduct that will be periodically re-examined and updated. This Declaration and any supplemental Guidelines and Codes will be placed on file with the International Academy of Astronautics (IAA) and made available on the IAA website.

**Principles**

1. **Handling Candidate Evidence:**

    - In the event of a putative detection of extraterrestrial intelligence, the discoverer should endeavour to make all efforts to authenticate and substantiate the detection, using the resources available to the discoverer and in collaboration with other investigators. Such efforts should ideally include, but not be limited to, independent observations or other examinations by multiple facilities and by more than one organisation utilising different instrumentation and methods.
    - Information about candidate signals or other potential detections should be handled with extreme care, recognising that initial findings may be incomplete or ambiguous, requiring thorough analysis and confirmation, which could be a lengthy process, and that follow-up study may not yield definitive conclusions. It is crucial to uphold the highest standards of scientific responsibility and integrity throughout this process, including recognition of the interests of humanity in the discovery. Best practices and tools in science communication should be employed to clearly convey the importance and significance of candidate discoveries to non-specialist audiences.

---

[1] 'Technosignatures' are defined as observable evidence of technology built or utilised by extraterrestrial beings e.g. narrow-band radio signals, laser emission, infrared excess associated with large-scale energy usage, anomalies in astronomical measurements due to megastructures etc, or an artefact. Technosignatures would indicate the presence of intelligent extraterrestrial life. This Declaration applies to the search for intelligent extraterrestrial life, not extraterrestrial life in general, nor to unidentified anomalous phenomena (UAP) in the Earth's atmosphere.

**2. Communicating and Sharing Information:**

DRAFT

- SETI practitioners and their institutions and organisations should be free to present reports on activities and results in public and professional fora. Individual practitioners shall have the right to decline from engaging directly or continually with the media, including social media, but shall use their best efforts to ensure that their organisation or institution provides updates on their science. Institutions and organisations should take appropriate steps for the safety of their researchers, and to protect them from negative professional repercussions.
- Institutions and organisations should be responsive to reasonable requests from news organisations, social media platforms, and other public communications media. Responses to inquiries should be prompt, accurate and honest.
- There is no obligation to disclose verification efforts until a discovery is confirmed. If a candidate technosignature is discovered, communication about ongoing observations and analyses may be necessary to dispel rumours and provide accurate and reliable information. Similarly, if analysis determines that a previously reported candidate technosignature is not extraterrestrial in origin, this should be promptly disclosed and clearly communicated.
- Speculative or unconfirmed statements and conclusions should be clearly identified as such.
- In their engagements with news and other media, institutions and organisations should provide accurate and timely information.

**3. Communicating Verification:**

- If the verification process satisfies Principle 1 and confirms – by the consensus of the other investigators involved and to a degree of certainty judged by the discoverers to be credible – that a signal or other evidence is due to extraterrestrial intelligence, the discoverers or their organisations or institutions should promptly report this conclusion in a full, complete and open manner to the public, the scientific community, and the Secretary General of the United Nations. The discoverers or their organisation or institution should have the privilege but not the obligation to make the first public announcement.
- The verification report should be peer-reviewed, and include the underlying data, the data analysis process and results of the verification efforts, any conclusions and interpretations, and any detected information content. This report should follow best practices in risk communication. The formal report should also be made to relevant organisations, including the International Academy of Astronautics, the International Astronomical Union, the Committee on Space Research (COSPAR) of the International Science Council, the International Institute of Space Law, the International Telecommunications Union, and the United Nations Committee on the Peaceful Uses of Outer Space, the Office of Outer Space Affairs, and other relevant U.N. bodies. Open access publication of verification data is encouraged.

**4. Monitoring, Archiving and Data Accessibility:**

- All data bearing on the evidence of extraterrestrial intelligence, together with the data analysis methods and code, should be preserved and disseminated to the international scientific community through refereed journal publications, meetings, conferences, websites appropriate for long-term archiving, and other appropriate means.
- The discovery should be continuously monitored, and best practices for the safe, reliable, and resilient handling of data should be employed. All data bearing on the evidence of extraterrestrial intelligence, including derived data products, should be recorded, and securely stored and archived to the greatest extent feasible and practicable, in at least two repositories in different geographic locations, and in a form that will make it accessible to observers and to the scientific community for replication of results and further analysis. The use of recognised international repositories and open standard formats is encouraged.

DRAFT

**5. Data and Frequency Protection:**

- Evidence of detection should be protected utilising best scientific practices, including tamper-proof records, and precautionary protocols. If the evidence of detection is in the form of electromagnetic signals, international agreement should be sought to protect the appropriate frequencies by exercising the extraordinary procedures established within the International Telecommunication Union.

**6. Post-Detection Protocol:**

- The IAA SETI Committee will maintain a Post-Detection Sub-Committee to assist and advise in matters that may arise in the event of a confirmed detection, and to support the scientific and public analysis by offering guidance, interpretation, and discussion of the wider implications of the discovery. This sub-committee should include international representation with science, ethics, law, social sciences, and communications professionals, as well as communications researchers with expertise in science and risk communication.
- The SETI Committee Executive Officers will support and assist the Post-Detection Sub-Committee with engagements with social media platforms and news organisations to responsively and effectively aid in the dissemination of accurate and reliable information.
- The IAA SETI Committee will collaborate with interdisciplinary researchers and working groups dedicated to Post-Detection issues and focus on subjects such as best practices for public communication about technosignature science.

**7. Communications with ETI following a confirmed detection:**

- SETI practitioners should cooperate with appropriate international consultations to consider whether a potential response to a confirmed detection of extraterrestrial intelligence should be made, and if so, its contents. Pending the outcome of such consultations, no reply should be sent. These consultations should be conducted through the United Nations and other broadly representative international bodies. The specific procedures for such consultations are to be outlined in a separate agreement, declaration, or arrangement to ensure a coordinated and responsible approach.[2]

**8. Ethical and Legal Considerations:**

- SETI practitioners shall adhere to the highest ethical standards, ensuring cooperation, honesty, and integrity in all aspects of their work. They will collaborate with international legal authorities to establish clear frameworks for managing the dissemination of information about potential extraterrestrial detections and comply with relevant laws and regulations. SETI practitioners shall cooperate in establishing and following ethical standards for the handling of any putative and/or confirmed detected evidence of extraterrestrial intelligence, including transparency and responsibility towards the global community.

This Updated Declaration replaces the previous Declaration of Principles Concerning the Conduct of the Search for Extraterrestrial Intelligence (SETI) adopted by the International Academy of Astronautics SETI Committee in 2010.

---

[2] This Declaration does not address the separate and distinct subject of messaging to extraterrestrial intelligence in advance of a confirmed detected extraterrestrial signal (METI).

**References**

[1] L. Walton, "The History of the IAA SETI Permanent Committee – 1980 to 1989" 70th International Astronautical Congress, Washington DC, IAC-19, A4,2,5, x52248

[2] L. Walton, "The History of the IAA SETI Permanent Committee – 1990 to 1999" 74th International Astronautical Congress (IAC), Baku, Azerbaijan, 2-6 October 2023. IAC-23-A4,2,1,x79493.

[3] "Declaration of Principles Concerning Activities Following the Detection of Extraterrestrial Intelligence" Adopted by the International Academy of Astronautics, 1989. https://iaaspace.org/wp-content/uploads/iaa/Scientific%20Activity/setideclaration.pdf (accessed 11 Sept 2025)

[4] "Draft Declaration of Principles Concerning Sending Communications with Extraterrestrial Intelligence", 1995. https://iaaspace.org/wp-content/uploads/iaa/Scientific%20Activity/setidraft.pdf (accessed 11 Sept 2025)

[5] M. Michaud, J. Billingham, J. Tarter, "A Reply From Earth" in *Acta Astronautica* Vol. 26, No. 3/4, pp. 295-297, 1992

[6] International Academy of Astronautics Position Paper: "A Decision Process for Examining the Possibility of Sending Communications to Extraterrestrial Civilizations: A Proposal", version 2007.http://resources.iaaseti.org/position.pdf (accessed 11 Sept 2025)

[7] IAA SETI Permanent Committee subcommittee on Communication with Extraterrestrial Intelligence, report by Ernst Fasan, Sept 2000: https://iaaseti.org/en/communication-extraterrestrial-intelligence-subcommittee-2000-ch/ (accessed 11 Sept 2025)

[8] United Nations Report of the Committee on the Peaceful Uses of Outer Space. United Nations, New York, 2000. General Assembly Official Records Fifty-fifth Session Supplement No. 20 (A/55/20) paragraphs 16, 157 www.unoosa.org/pdf/gadocs/A_55_20E.pdf (accessed 11 Sept 2025)

[9] "Declaration of Principles Concerning the Conduct of the Search for Extraterrestrial Intelligence", IAA SETI Permanent Study Group, 2010. https://iaaspace.org/wp-content/uploads/iaa/Scientific%20Activity/setideclaration2.pdf (accessed 11 Sept 2025)

[10] C. Oliver, K. Denning, L. I. Tennen., M. Garrett, "Revising the SETI Post-Detection Protocols for the 2020s and Beyond: A Report on Work in Progress", IAC-23-A4.2.3 x79618, 74th International Astronautical Congress (IAC), Baku, Azerbaijan, 2023. IAFAstro archive.

[11] L. I. Tennen, K. Denning, C. Oliver, M. Garrett, "The Future of the SETI Post-Detection Protocols: Progress Towards Revisions", IAC-24-A4.2.6 x84417, 75th International Astronautical Congress (IAC), Milan, Italy, 2024. IAFAstro archive.

[12] J. Bimm, M. Voytek, C. Scharf et al. "Communicating Discoveries in the Search for Life in the Universe (CDSLU) Workshop Report" Forthcoming in *Astrobiology* Volume 25, No. 11, Nov 2025, and on ArXiv.

[13] NASA Decadal Astrobiology Research and Exploration Strategy, https://science.nasa.gov/astrobiology/strategy/dares/ (accessed 11 Sept 2025)

[14] L. I. Tennen, K. Denning, C. Oliver, M. Garrett, "The Future of the SETI Post-Detection Protocols: Progress Towards Revisions", IAC-24-A4.2.6 x84417, 75th International Astronautical Congress (IAC), Milan, Italy, 2024. IAFAstro archive.